\documentclass[12pt]{article}

\usepackage{amsmath,amsfonts,latexsym,amssymb,amscd}
\usepackage{pslatex}
\usepackage[latin1]{inputenc}
\usepackage[T1]{fontenc}
\usepackage{pspicture}
\usepackage{verbatim,amsthm,curves,graphics}
\usepackage{mathrsfs}

\usepackage{graphicx}

\begin{document}


\title{ Horizon area-angular momentum-charge-magnetic fluxes inequalities in 5D Einstein-Maxwell-dilaton gravity  }

\author{
Stoytcho Yazadjiev$^{}$\thanks{\tt yazad@phys.uni-sofia.bg}
\\ \\
{\it $ $Department of Theoretical Physics, Faculty of Physics, Sofia
University} \\
{\it 5 J. Bourchier Blvd., Sofia 1164, Bulgaria} \\
    }
\date{}

\maketitle

\begin{abstract}
In the present paper we consider 5D spacetimes satisfying the Einstein-Maxwell-dilaton gravity
equations which are  $U(1)^2$ axisymmetric but otherwise highly dynamical.
We derive inequalities between the area, the angular momenta, the electric charge  and
the magnetic fluxes for any smooth stably outer marginally trapped surface.
\end{abstract}


\sloppy

\section{Basic notions and setting the problem}

The study of inequalities between the horizon area and the other
characteristics of the horizon has attracted a lot of interest
recently. Within the general theory of relativity, lower bounds for
the area of dynamical  horizons in terms of their angular momentum
or/and charge were given in \cite{ADC}--\cite{CJR}, generalizing the
similar inequalities for the stationary black holes
\cite{AHC}--\cite{HCA}. These remarkable inequalities are based
solely on general assumptions and they hold for any axisymmetric but
otherwise highly dynamical horizon in general relativity. For a nice
review on the subject we refer the reader to \cite{D}.
The relationship between the proofs of the area-angular-momentum-charge
inequalities for quasilocal black holes and stationary black holes
is discussed in \cite{CENS}-\cite{J}.
Inequalities between the horizon  area, the angular momentum,  and the charges were also
studied in some 4D alternative gravitational theories \cite{Y1}.

A generalization of the 4D horizon area-angular momentum inequality  to
D-dimensional vacuum Einstein gravity with $U(1)^{D-3}$ group of
spatial isometries was given in \cite{H}. The purpose of the present
work is to derive some inequalities between the horizon area,
horizon angular momentum, horizon charges and magnetic fluxes in the 5D
Einstein-Maxwell-dilaton gravity including as a particular case the
5D Einstein-Maxwell gravity.  It should be stressed that the
derivation of the  mentioned inequalities in the higher dimensional
Einstein-Maxwell and Einstein-Maxwell-dilaton gravity is much more
difficult and is not so straightforward as in the higher dimensional
vacuum gravity even in spacetimes admitting $U(1)^{D-3}$ isometry
group. The main reason behind this is the lack of nontrivial group
of hidden symmetries for the dimensionally reduced
Einstein-Maxwell-dilaton equations in the general case \footnote{Fortunately, there are sectors in Einstein-Maxwell-gravity which are completely integrable
 \cite{Y3}-\cite{Y5}.}\cite{Y2}. In contrast, the dimensionally reduced vacuum Einstein
equations (in spacetimes with $U(1)^{D-3}$ isometry group) possess
nontrivial group of hidden symmetries, namely $SL(D-2,\mathbb{R})$ and a
matrix sigma model presentation is possible. Some of the
difficulties due to the presence of a Maxwell field can be
circumvented by following a method similar to that used in the 4D
Einstein-Maxwell-dilaton gravity \cite{Y1} as we show below.

Let $({\cal M},g_{ab},F_{ab},\varphi)$ be  a $5$-dimensional  spacetime satisfying
the Einstein-Maxwell-dilaton equations
\begin{eqnarray}\label{FE}
&&G_{ab}= 2\partial_a\varphi \partial_b\varphi  - \nabla^{c}\varphi\nabla_{c}\varphi g_{ab} - 2V(\varphi)g_{ab} +
 2 e^{-2\alpha\varphi}\left(F_{ac}F_{b}{}^c- \frac{g_{ab}}{4}F_{cd}F^{cd} \right),\\
&&\nabla_{a}\left( e^{-2\alpha\varphi }F^{ab}\right)=0=\nabla_{[a} F_{bc]}, \\
&&\nabla_{a}\nabla^a \varphi = \frac{dV(\varphi)}{d\varphi}- \frac{\alpha}{2}
e^{-2\alpha\varphi} F_{cd}F^{cd} ,
\end{eqnarray}
where $g_{ab}$ is the spacetime metric, $\nabla_a$ is its
Levi-Civita connection,  $G_{ab}=R_{ab}- \frac{1}{2}g_{ab}R$ is
the Einstein tensor and $F_{ab}$ is the Maxwell field. The dilaton field is denoted by
$\varphi$, $V(\varphi)$ is its potential and $\alpha$ is the dilaton coupling parameter.
We assume that the dilaton potential is non-negative, $V(\varphi)\ge 0$.
The Einstein-Maxwell gravity is recovered by first putting
$\alpha=0$ and $V(\varphi)=0$ and then $\varphi=0$.

As an additional technical assumption we require  the spacetime to admit $U(1)^2$ group of spatial isomerties.
The commuting Killing fields are denoted by $\eta_1$ and $\eta_2$ and they are normalized to have a period $2\pi$.
We also require the Maxwell and the dilaton fields to be invariant under the flow of the Killing fields, i.e.
$\pounds_{\eta_{I}} F=\pounds_{\eta_{I}} \varphi=0$.

Let us further consider a compact closed smooth   submanifold ${\cal B}$ of dimension $\dim {\cal B}= 3$ invariant under the action of $U(1)^2$.
The induced metric on ${\cal B}$ and its Levi-Civita connection are denoted by $\gamma_{ab}$ and $D_{a}$, respectively.
The future directed null normals to ${\cal B}$ will be denoted by $n$ and $l$ with the normalization condition $g(n,l)=-1$ and
with $-l$ pointing outward. In what follows we require ${\cal B}$ to be a stably outer marginally trapped surface which means that
$\Theta_{n}=0$ and $\pounds_{l}\Theta_{n}\le 0$ with $\Theta_{n}$ being the expansion of $n$ on ${\cal B}$.

As a 3-dimensional compact manifold with an action of $U(1)^2$, $\cal B$ is topologically either $S^3$, $S^2\times S^1$ or a lens space $L(p,q)$ with
$p$ and $q$ being  co-prime integers \cite{HY1,HY2}. Moreover, the factor space ${\hat {\cal B}}={\cal B}/U(1)^2$ can be identified with the closed interval $[-1,+1]$.
As it was shown in \cite{HY1,HY2},  certain linear combinations of the Killing fields $\eta_{I}$, with integer coefficients, vanish at the ends of the
factor space. In other words, there exist integer vectors ${\bf a}_{\pm }\in \mathbb{Z}^2$ such that  $a^{I}_{\pm}\eta_{I}\to 0$ at $x=\pm 1$,
where $x$ is the coordinate
parameterizing the factor space. Equivalently, the Gram matrix defined by
\begin{eqnarray}
H_{IJ}=g(\eta_I,\eta_J)
\end{eqnarray}
is invertible in the interior of the interval $[-1,1]$ and has one-dimensional kernel at the interval end points, i.e. $H_{IJ}a^{I}_{\pm}\to 0$ at
$x=\pm 1$.

In fact, the integer vectors ${\bf a}_{\pm}$ determine the topology of ${\cal B}$. By a global $SL(2,\mathbb{Z})$ redefinition of the Killing fields \cite{HY1,HY2}
we may present ${\bf a}_{\pm}$ in the form ${\bf a}_{+}=(1,0)$ and ${\bf a}_{-}=(p,q)$ with $p$ and $q$ being coprime integers. The topology of ${\cal B}$
is then $S^3$ when $(p=\pm 1, q=0)$, $S^2\times S^1$ when $(p=0, q=\pm 1)$ and that of a lens space $L(p,q)$ in the other cases.

Proceeding further we consider a small neighborhood $O$ of ${\cal B}$. When the neighborhood is sufficiently small it can be foliated by two-parametric copies
 ${\cal B}(u,r)$ of ${\cal B}={\cal B}(0,0)$ and parameterized by the so-called null Gauss coordinates defined by a well-known procedure \cite{HIW}.
In Gauss null coordinates the metric in $O$ can be written in the form

\begin{eqnarray}\label{metric}
g= - 2du(dr - r^2 \Upsilon \,du - r \beta_{a}dy^a) + \gamma_{ab}dy^ady^b ,
\end{eqnarray}
where $n=\frac{\partial}{\partial u}$, $l=\frac{\partial}{\partial r}$, and the function $\Upsilon$
and the metric $\gamma$ are invariantly defined on each ${\cal B}(u,r)$. Using these coordinates one can show that on ${\cal B}$
it holds

\begin{eqnarray}\label{INE1}
R_{\gamma} - D^a\beta_a - \frac{1}{2}\beta^a\beta_a - 2G_{ab}n^al^b= -2{\pounds_{l}\Theta_{n}}\ge 0 ,
\end{eqnarray}
where it has been taken into account that $\Theta_{n}=0$ on ${\cal B}$. Here $R_{\gamma}$  and $D^a$ are the Ricci scalar curvature and
Levi-Civita connection with respect to the metric  $\gamma_{ab}$ on ${\cal B}$. Taking into account that the dilaton potential is nonnegative
this inequality can be rewritten in the form

\begin{eqnarray}\label{INE2}
R_{\gamma} - D^a\beta_a - \frac{1}{2}\beta^a\beta_a - 2{\tilde G}_{ab}n^al^b= 2V(\varphi) -2{\pounds_{l}\Theta_{n}}\ge 0 ,
\end{eqnarray}
where ${\tilde G}_{ab}= G_{ab} + 2V(\varphi)g_{ab}$.

Making use of (\ref{INE2}), for every axisymmetric function
$f$ (i.e. every function $f$ invariant under the isometry group) we have

\begin{eqnarray}
0\le\int_{{\cal B}} \left( - D^a\beta_a - \frac{1}{2}\beta^a\beta_a + R_{\gamma}- 2{\tilde G}_{ab}n^al^b\right)f^2 dS  \nonumber\\ =
\int_{{\cal B}} \left( 2 f\beta^a D_a f - \frac{1}{2}\beta^a\beta_a f^2 + R_{\gamma}f^2 - 2{\tilde G}_{ab}n^al^b f^2\right) dS ,
\end{eqnarray}
where $dS$ is the surface element on ${\cal B}$. Now we consider the unit tangent vector $N^a$ on ${\cal B}$ which is orthogonal to
$\eta_{I}$. With its help and taking into account that $\left(\gamma^{ab} - N^aN^b\right)D_{b}f=0$,
we find $\beta_a\beta^a= \left(\gamma^{ab} - N^aN^b\right)\beta_a\beta_b + \left(N^a\beta_a\right)^2$ and
$2f\beta^aD_{a}f= 2\left(fN^a\beta_a\right) \left( N^bD_{b}f\right)$ which gives

\begin{eqnarray}
0\le \int_{\cal B} \left[ 2\left(fN^a\beta_a\right) \left( N^bD_{b}f\right) -\frac{1}{2} \left(N^a\beta_a\right)^2 f^2
- \frac{1}{2}\left(\gamma^{ab} - N^aN^b\right)\beta_{a}\beta_{b}f^2 +
    R_{\gamma}f^2 - 2{\tilde G}_{ab}n^al^b f^2 \right]dS .
\end{eqnarray}

Finally, taking into account that

\begin{eqnarray}
2\left(fN^a\beta_a\right) \left( N^bD_{b}f\right) -\frac{1}{2} \left(N^a\beta_a\right)^2 f^2 \le 2\left(N^bD_bf\right)^2
\end{eqnarray}
and that $\left(N^bD_bf\right)^2= \left(\gamma^{ab} - N^aN^b\right)D_{a}fD_bf + N^aN^bD_afD_bf=\gamma^{ab}D_{a}fD_bf = D_afD^af$
we obtain the important inequality

\begin{eqnarray}\label{INEQ}
0\le \int_{{\cal B}} \left[2D_afD^af - \frac{1}{2}\left(\gamma^{ab} - N^aN^b\right)\beta_{a}\beta_{b}f^2 + R_{\gamma}f^2 - 2{\tilde G}_{ab}n^al^b f^2 \right]dS .
\end{eqnarray}

In order to extract the constructive information from this inequality we should perform a dimensional reduction and  express the inequality
as an inequality on the factor space ${\hat {\cal B}}={\cal B}/U^2(1)=[-1,1]$. The dimensional reduction can be performed along the lines of \cite{Y2}. So we shall
give here only some basic steps and results without going into detail. As a first step it is very convenient to present the Killing fields $\eta_{I}$
in adapted coordinates, i.e. $\eta_{I}=\frac{\partial}{\partial \phi^{I}}$ where the coordinates $\phi^{I}$  are $2\pi$-periodic.
Then the induced metric $\gamma_{ab}$ on ${\cal B}$ takes the form

\begin{eqnarray}\label{metric_gamma}
\gamma= \frac{dx^2}{C^2h} + H_{IJ}d\phi^{I}d\phi^{J} ,
\end{eqnarray}
where $C>0$ is a constant and $h=\det(H_{IJ})$. The absence of conical singularities requires the following condition to be satisfied

\begin{eqnarray}\label{conical_con}
\lim_{x\rightarrow \pm 1} C^2\frac{h}{1-x^2} \frac{H_{IJ}a^I_{\pm}a^{J}_{\pm}}{1-x^2}=1.
\end{eqnarray}
The area ${\cal A}$ of ${\cal B}$ can be easily found from (\ref{metric_gamma}) and the result is

\begin{eqnarray}
{\cal A}= 8\pi^2 C^{-1}.
\end{eqnarray}
Therefore the condition (\ref{conical_con}) can be rewritten in the form

\begin{eqnarray}\label{A_conical}
\frac{{\cal A}}{8\pi^2} = \lim_{x\rightarrow  1}\left(\frac{h}{1-x^2} \frac{H_{IJ}a^I_{+}a^{J}_{+}}{1-x^2}\right)^{1/4}
\lim_{x\rightarrow  -1}\left(\frac{h}{1-x^2} \frac{H_{IJ}a^I_{-}a^{J}_{-}}{1-x^2}\right)^{1/4} .
\end{eqnarray}

Since the factor space $O/U(1)^2$ is simply connected we can introduce electromagnetic potentials $\Phi_{I}$ and $\Psi$
invariant under the isometry group and defined by

\begin{eqnarray}
d\Phi_{I}=i_{\eta_{I}}F, \;\;\; d\Psi= e^{-2\alpha\varphi} i_{\eta_2}i_{\eta_1}\star F .
\end{eqnarray}

The Maxwell 2-form can then be  written in the form

\begin{eqnarray}
F= H^{IJ}\eta_{I}\wedge d\Phi_{J} + h^{-1} e^{2\alpha\varphi}\star \left(d\Psi\wedge\eta_1\wedge \eta_2 \right).
\end{eqnarray}

Using the field equations one can show that there exist potentials $\chi_{I}$ invariant under the isometry group such that
the twist $\omega_{I}=\star \left(\eta_{1}\wedge \eta_2 \wedge d\eta_{I}\right)$ satisfies

\begin{eqnarray}
\omega_{I}= d\chi_{I} + 2\Phi_{I}d\Psi - 2\Psi d\Phi_{I}.
\end{eqnarray}
By direct computation of the twist using the metric (\ref{metric}) one finds that on ${\cal B}$ it holds
$\beta_{I}=i_{\eta_{I}}\beta=Ci_{\frac{\partial}{\partial x}}\omega_{I}$ or
in explicit form

\begin{eqnarray}\label{beta_I}
\beta_{I}= \partial_{x}\chi_{I} +  2\Phi_{I}\partial_{x}\Psi - 2\Psi \partial_{x}\Phi_{I} .
\end{eqnarray}

Also one can show that on ${\cal B}$ we have

\begin{eqnarray}\label{G_AB}
{\tilde G}_{ab}n^al^b = D_a\varphi D^a\varphi + 2e^{-2\alpha\varphi} H^{IJ}D_{a}\Phi_{I}D^a\Phi_{J} + h^{-1}e^{2\alpha\varphi}D_{a}\Psi D^a\Psi .
\end{eqnarray}

Using the explicit form (\ref{metric_gamma}) of the metric induced  on ${\cal B}$, by direct computation we find

\begin{eqnarray}\label{R_gamma}
R_{\gamma}= C^2 h \left[-\frac{\partial^2_{x}h}{h}  +  \frac{1}{4}h^{-2}(\partial_{x}h)^2 - \frac{1}{4}Tr\left(H^{-1}\partial_{x}H\right)^2   \right].
\end{eqnarray}

Finally, choosing

\begin{eqnarray}
f= \left(\frac{1-x^2}{h}\right)^{1/2},
\end{eqnarray}
substituting (\ref{beta_I}), (\ref{G_AB}) and (\ref{R_gamma}) into (\ref{INEQ}) and taking into account that $dS= C^{-1}dx\prod_{I}d\phi^{I}$, we obtain

\begin{eqnarray}\label{INEQ1}
&&\int^{1}_{-1} \left\{ (1-x^2)\left[\frac{1}{8} Tr\left(H^{-1}\partial_{x}H\right)^2 + \frac{1}{8}h^{-2}(\partial_{x}h)^2 +
 \right.\right.  \nonumber \\ &&\left. \left.
\frac{1}{4}h^{-1}H^{IJ}\left(\partial_{x}\chi_{I} +  2\Phi_{I}\partial_{x}\Psi - 2\Psi \partial_{x}\Phi_{I}\right) \left(\partial_{x}\chi_{J}
+  2\Phi_{J}\partial_{x}\Psi - 2\Psi \partial_{x}\Phi_{J}\right)   \right.\right.  \nonumber \\ &&\left. \left.
+ e^{-2\alpha\varphi }H^{IJ}\partial_{x}\Phi_{I}\partial_{x}\Phi_{J} + e^{2\alpha\varphi} h^{-1}(\partial_x\Psi)^2 + (\partial_{x}\varphi)^2   \right]
- \frac{1}{1-x^2} \right\}dx \le 0.
\end{eqnarray}

Now we can introduce the strictly positive definite metric $G_{AB}$ given by

\begin{eqnarray}
&&G_{AB} dX^A dX^B = \frac{1}{8} Tr\left(H^{-1}dH\right)^2 + \frac{1}{8}h^{-2}(dh)^2 + \nonumber \\
&&\frac{1}{4}h^{-1}H^{IJ}\left(d\chi_{I} +  2\Phi_{I}d\Psi - 2\Psi d\Phi_{I}\right) \left(d\chi_{J}
+ 2\Phi_{J}d\Psi - 2\Psi d\Phi_{J}\right) + \nonumber \\ &&  e^{-2\alpha\varphi }H^{IJ}d\Phi_{I}d\Phi_{J} + e^{2\alpha\varphi} h^{-1}(d\Psi)^2 + (d\varphi)^2
\end{eqnarray}
on the 9-dimensional manifold ${\cal N}=\{(H_{IJ}(I\le J), \chi_{I}, \Phi_{I}, \Psi,\phi)\in \mathbb{R}^9; h>0 \}$. In terms of this metric the
inequality (\ref{INEQ1}) takes the form

\begin{eqnarray}\label{I_INEQ}
I_{*}[X^A]=\int^{1}_{-1}\left[(1-x^2)G_{AB}\frac{dX^A}{dx}\frac{dX^B}{dx} - \frac{1}{1-x^2}\right]dx \le 0 .
\end{eqnarray}

In order to transform this inequality into an inequality for the area, we use condition (\ref{A_conical}) which combined with ({\ref{I_INEQ}})
gives

\begin{eqnarray}\label{A_INEQ}
{\cal A}\ge 8\pi^2 e^{I[X^A]},
\end{eqnarray}
where

\begin{eqnarray}\label{I_FUN}
I[X^A]= I_{*}[X^A] + \frac{1}{4}x\ln\left[\frac{h}{1-x^2} \frac{H_{IJ}a^I(x)a^J(x)}{1-x^2}\right]|^{x=1}_{x=-1}
\end{eqnarray}
with $a^{I}(x)$ defined by $a^I(x)=\frac{1}{2}(1+x)a^{I}_{+} + \frac{1}{2}(1-x)a^{I}_{-}$. We should note that there is an ambiguity in
defining the functional $I[X^A]$. For example, we can define it  by

\begin{eqnarray}
I[X^A]= a I_{*}[X^A] + \frac{1}{4}x\ln\left[\frac{h}{1-x^2} \frac{H_{IJ}a^I(x)a^J(x)}{1-x^2}\right]|^{x=1}_{x=-1},
\end{eqnarray}
where $a$ is an arbitrary positive number. This ambiguity, however, does not affect the final results since $I_{*}[X^A]=0$
as we show below.

\section{Minimizer existence lemma}

In order to put a lower bound on the area we should find the minimum of the function $I[X^A]$ with appropriate boundary conditions
if the minimum exists.  Below we show that in certain cases the minimum  exists. The natural class of functions for the minimizing problem
is given by $\sigma =-\ln\left(\frac{h}{1-x^2}\right)\in C^{\infty}[-1,1]$, $\ln\left(\frac{H_{IJ}a^{I}a^{J}}{1-x^2}\right)\in C^{\infty}[-1,1]$,
$(\chi_{I},\Phi_{I},\Psi,\varphi)\in C^{\infty}[-1,1]$ with boundary conditions $\sigma(\pm 1)=\sigma^{\pm}$, $(\chi_{I}(\pm 1),\Phi_{I}(\pm 1),\Psi(\pm 1),
\varphi(\pm 1))= (\chi^{\pm}_{I},\Phi^{\pm}_{I},\Psi^{\pm},\varphi^{\pm})$. Since the electromagnetic potentials and the twist potential are defined up to a constant,
without loss of generality we can choose

\begin{eqnarray}
\chi^{+}_{I}= -\chi^{-}_{I}, \; \Phi^{+}_{I}= -\Phi^{-}_{I}, \; \Psi^{+}= -\Psi^{-} .
\end{eqnarray}

\medskip
\noindent {\bf Lemma 1.} {\it For dilaton coupling parameter
satisfying  $0\le\gamma^2\le \frac{8}{3}$, there exists a unique smooth
minimizer of the functional $I[X^A]$
with the prescribed boundary conditions.}

\medskip
\noindent
{\bf Proof.}  Let us consider the truncated functional

\begin{eqnarray}
I_{*}[X^A][x_2,x_1]=\int^{x_2}_{x_1} \left[(1-x^2) G_{AB}\frac{dX^A}{dx}\frac{dX^{B}}{dx}  - \frac{1}{1-x^2}\right]dx
\end{eqnarray}
with $-1<x_1< x_2<1$. By introducing  a new variable $t=\frac{1}{2}\ln\left(\frac{1+x}{1-x}\right)$ the truncated functional takes the form

\begin{eqnarray}
I_{*}[X^A][t_2,t_1]=\int^{t_2}_{t_1} \left[G_{AB}\frac{dX^A}{dt}\frac{dX^{B}}{dt} - 1\right]dt,
\end{eqnarray}
which is just a modified version of the geodesic functional in the Riemannian space $({\cal N}, G_{AB})$. Consequently the critical points of
the functional  are geodesics  in ${\cal N}$. It was shown in \cite{Y2} that for $0\le\alpha^2 \le \frac{8}{3} $ the Riemannian space $({\cal N}, G_{AB})$
is simply connected, geodesically complete and with negative sectional curvature. Therefore, for fixed points $X^A(t_1)$ and $X^a(t_2)$ there exist a unique minimizing
geodesic connecting these points. Therefore the global minimizer of $I_{*}[X^A][t_2,t_1]$ exists and is unique for $0\le\alpha^2 \le \frac{8}{3} $.
Since $({\cal N}, G_{AB})$ is geodesically  complete the global minimizer of $I_{*}[X^A][t_2,t_1]$ can be extended to a global minimizer of
$I_{*}[X^A]$. Indeed, let us take  $x_{1}(\epsilon)=-1 + \epsilon$ and $x_{2}(\epsilon)=1-\epsilon$
(i.e. $t_1(\epsilon)=-t_2(\epsilon)=\frac{1}{2}\ln\left(\frac{\epsilon}{2-\epsilon}\right)$) with $\epsilon$ being a small positive number
and consider the truncated functional

\begin{eqnarray}
I_{*}^{\,\epsilon}[X^A]= \int^{x_2(\epsilon)}_{x_1(\epsilon)} \left[(1-x^2) G_{AB}\frac{dX^A}{dx}\frac{dX^{B}}{dx}  - \frac{1}{1-x^2}\right]dx
\end{eqnarray}
with boundary conditions $X^A(x_1(\epsilon))$ and $X^A(x_2(\epsilon))$.  Consider now the unique minimizing geodesic $\Gamma_{\epsilon}$ in ${\cal N}$
between the points $X^A(x_1(\epsilon))$ and $X^A(x_2(\epsilon))$. Then we have

\begin{eqnarray}\label{INQ_FUNC}
I_{*}^{\,\epsilon}[X^A]\ge I_{*}^{\,\epsilon}[X^A]|_{\Gamma_{\epsilon}}
\end{eqnarray}
where the right hand side of the above inequality is evaluated on the geodesic $\Gamma_{\epsilon}$. Taking into account that
$\lambda^2_{\epsilon}=G_{AB}\frac{dX^A}{dt}\frac{dX^B}{dt}$ is constant on the geodesic $\Gamma_{\epsilon}$, we obtain

\begin{eqnarray}
I_{*}^{\,\epsilon}[X^A]|_{\Gamma_{\epsilon}}= \int^{t_2(\epsilon)}_{t_1(\epsilon)} \left[G_{AB}\frac{dX^A}{dt}\frac{dX^{B}}{dt} - 1\right]dt=
\left(\lambda^2_{\epsilon} -1\right) \left(t_2(\epsilon) - t_1(\epsilon)\right).
\end{eqnarray}

Our next step is to evaluate $\lambda^2_{\epsilon}$ and this can be done by evaluating $G_{AB}\frac{dX^A}{dt}\frac{dX^B}{dt}$ at the
boundary points which are in a small neighborhood of the poles $x=\pm 1$. For this purpose we first write $\lambda^2_{\epsilon}$ in the form

\begin{eqnarray}
\lambda^2_{\epsilon} = &&\frac{(1-x^2)^2}{8} Tr\left(H^{-1}\frac{dH}{dx}\right)^2 + \frac{(1-x^2)^2}{8}h^{-2}\left(\frac{dh}{dx}\right)^2
 \\ &&  +
\frac{(1-x^2)^2}{4}h^{-1}H^{IJ}\left(\frac{d\chi_{I}}{dx} +  2\Phi_{I}\frac{d\Psi}{dx} - 2\Psi \frac{d\Phi_{I}}{dx}\right) \left(\frac{d\chi_{J}}{dx}
+ 2\Phi_{J}\frac{d\Psi}{dx} - 2\Psi \frac{d\Phi_{J}}{dx}\right)  \nonumber \\ && +  (1-x^2)^2 e^{-2\alpha\varphi }H^{IJ}\frac{d\Phi_{I}}{dx}\frac{d\Phi_{J}}{dx}
+ (1-x^2)^2e^{2\alpha\varphi} h^{-1}\left(\frac{d\Psi}{dx}\right)^2 + (1-x^2)^2\left(\frac{d\varphi}{dx}\right)^2 .\nonumber
\end{eqnarray}

Within the class of functions we consider, we have

\begin{eqnarray}
\frac{(1-x^2)^2}{8}h^{-2}\left(\frac{dh}{dx}\right)^2= \frac{1}{2} + O(\epsilon)
\end{eqnarray}
in a small neighborhood of the poles.

In order to estimate the term associated with $H$ we take into account that $H^{-1}\frac{dH}{dx}$ satisfies its own characteristic equation, namely
$Tr\left(H^{-1}\frac{dH}{dx}\right)^2= h^{-2}\left(\frac{dh}{dx}\right)^2 - 2h^{-1}\det \frac{dH}{dx}$. Hence we find

\begin{eqnarray}
\frac{(1-x^2)^2}{8}Tr\left(H^{-1}\frac{dH}{dx}\right)^2= \frac{1}{2} + O(\epsilon).
\end{eqnarray}

Proceeding further we notice that $\partial/\partial \chi_{I}$ are Killing fields for the metric $G_{AB}$ and consequently we have the following
constants of motion on the geodesics $\Gamma_{\epsilon}$

\begin{eqnarray}
\frac{1}{2}h^{-1} H^{IJ} \left(\frac{d\chi_{I}}{dt} +  2\Phi_{I}\frac{d\Psi}{dt} - 2\Psi \frac{d\Phi_{I}}{dt}\right)=
\frac{1-x^2}{2}h^{-1} H^{IJ} \left(\frac{d\chi_{I}}{dx} +  2\Phi_{I}\frac{d\Psi}{dx} - 2\Psi \frac{d\Phi_{I}}{dx}\right)=c^{I}_{\epsilon}.\nonumber \\
\end{eqnarray}
Hence we obtain

\begin{eqnarray}
&&\frac{(1-x^2)^2}{4}h^{-1}H^{IJ}\left(\frac{d\chi_{I}}{dx} +  2\Phi_{I}\frac{d\Psi}{dx} - 2\Psi \frac{d\Phi_{I}}{dx}\right) \left(\frac{d\chi_{J}}{dx}
+ 2\Phi_{J}\frac{d\Psi}{dx} - 2\Psi \frac{d\Phi_{J}}{dx}\right) = \nonumber \\
&& h H_{IJ}\,\, c^{I}_{\epsilon} \, c^{J}_{\epsilon}= O(\epsilon).
\end{eqnarray}
For the remaining terms, it is easy to see that they behave as

\begin{eqnarray}
&&(1-x^2)^2 e^{-2\alpha\varphi }H^{IJ}\frac{d\Phi_{I}}{dx}\frac{d\Phi_{J}}{dx}= O(\epsilon), \\
&&(1-x^2)^2e^{2\alpha\varphi} h^{-1}\left(\frac{d\Psi}{dx}\right)^2= O(\epsilon), \\
&&(1-x^2)^2\left(\frac{d\varphi}{dx}\right)^2 = O(\epsilon^2).
\end{eqnarray}

Summarizing the results so far, we conclude that the behavior of $\lambda^2_{\epsilon}$ for small $\epsilon$ is

\begin{eqnarray}
\lambda^{2}_{\epsilon}= 1 + O(\epsilon).
\end{eqnarray}
Therefore we have

\begin{eqnarray}
\lim_{\epsilon \to 0} I_{*}^{\,\epsilon}[X^A]|_{\Gamma_{\epsilon}}=0
\end{eqnarray}
which, in view  of (\ref{INQ_FUNC}), gives

\begin{eqnarray}\label{I_INEQ_PLUS}
I_{*}[X^A]=\lim_{\epsilon\to 0}I_{*}^{\,\epsilon}[X^A]\ge 0.
\end{eqnarray}

Therefore, there exists a  unique global minimizer of  the functional $I_{*}[X^A]$. Since the functionals  $I[X^A]$ and $I_{*}[X^A]$ differ in boundary terms
the global minimizer of $I_{*}[X^A]$  is also a global minimizer of $I[X^A]$. This completes the proof.

It should be noted that from (\ref{I_INEQ}) and (\ref{I_INEQ_PLUS}) immediately follows that $I_{*}[X^A]=0$. 

The extremal stationary near horizon geometry is in fact defined by the same variational problem with the same boundary conditions and by the same class of functions.
Therefore, as an direct consequence of the proven lemma we obtain the following

\medskip
\noindent {\bf Corollary.} {\it For every dilaton coupling parameter $\alpha$ in the range  $0\le \alpha^2 \le \frac{8}{3}$ the area ${\cal A}$ of $\cal B$
satisfies the inequality}
\begin{eqnarray}
{\cal A}\ge {\cal A}_{ENHG},
\end{eqnarray}
{\it where ${\cal A}_{ENHG}$ is the area associated with the extremal stationary near horizon geometry of
Einstein-Maxwell-dilaton gravity with $V(\varphi)=0$, for the corresponding $\alpha$.
The equality is saturated only for the area associated with extremal stationary near horizon geometry with $V(\varphi)=0$.}

\section{Horizon area-angular momenta-charge-magnetic fluxes inequality for critical dilaton coupling parameter}

For the critical coupling $\alpha^2=\frac{8}{3}$ the Riemannian space $({\cal N}, G_{AB})$ is an $SL(4, \mathbb{R})/O(4)$ symmetric space \cite{Y1}
and therefore, there exists a matrix $M$ such that the metric $G_{AB}$ can be written in the form

\begin{eqnarray}
G_{AB}dX^A dX^B = \frac{1}{8} Tr \left(M^{-1}dM \right)^2 ,
\end{eqnarray}
where $M$ is positive definite and $M\in SL(4,\mathbb{R})$. Finding the explicit form of the matrix $M$ is a tedious task and here we present only the final result.
The matrix $M$ is given by

\begin{eqnarray}
M = \left(
      \begin{array}{cc}
        E_{2\times 2} & 0 \\
        S^T & E_{2\times 2}  \\
      \end{array}
    \right) \left(
              \begin{array}{cc}
                N & 0 \\
                0 & Y \\
              \end{array}
            \right) \left(
      \begin{array}{cc}
        E_{2\times 2} & S \\
        0 & E_{2\times 2}  \\
      \end{array}
    \right)= \left(
               \begin{array}{cc}
                 N & NS \\
                 S^{T}N &  S^{T}NS + Y \\
               \end{array}
             \right) ,
\end{eqnarray}
where $E_{2\times 2}$ is the unit $2\times 2$ matrix and $S$,  $N$ and $Y$ are $2\times 2$ matrices which have the following explicit form:
\begin{eqnarray}
S= \left(
     \begin{array}{cc}
       2\Phi_1 & 2\Phi_2 \\
       \chi_1 + 2\Phi_1\Psi & \chi_2 + 2\Phi_2\Psi \\
     \end{array}
   \right) ,
\end{eqnarray}

\begin{eqnarray}
N= e^{\sqrt{\frac{2}{3}}\,\varphi} h^{-1}\left(
                                     \begin{array}{cc}
                                       e^{-4\sqrt{\frac{2}{3}}\,\varphi}h  + 4\Psi^2 & -2\Psi  \\
                                       -2\Psi  & 1 \\
                                     \end{array}
                                   \right),
\end{eqnarray}

\begin{eqnarray}
Y= e^{\sqrt{\frac{2}{3}}\,\varphi} H .
\end{eqnarray}

In terms of the matrix $M$, the Euler-Lagrange equations are

\begin{eqnarray}
\frac{d}{dx}\left[(1-x^2)M^{-1}\frac{dM}{dx}\right]=0.
\end{eqnarray}
Hence we obtain

\begin{eqnarray}
(1-x^2)M^{-1}\frac{dM}{dx}= 2A ,
\end{eqnarray}
where $A$ is a constant matrix with $Tr A=0$, since $\det M=1$. Integrating further we find

\begin{eqnarray}
M= M_{0}\exp{\left(\ln\frac{1+x}{1-x}A\right)}
\end{eqnarray}
with $M_{0}$ being a constant matrix with the same properties as $M$ and satisfying $A^{T}M_{0}=M_{0}A$. As a positive definite matrix, $M_{0}$
can be written in the form $M_{0}=BB^{T}$ for some constant matrix $B$ with $|\det B |=1$ and this presentation is up to an orthogonal matrix  $O$, i.e
it is invariant under the transformation $B\longrightarrow BO$. This freedom can be used to diagonalize the symmetric matrix
$B^{T}A {B^{T}}^{-1}$. So we can take $B^{T}A {B^{T}}^{-1}=diag(\lambda_1, \lambda_2, \lambda_3, \lambda_4)$ and we obtain

\begin{eqnarray}\label{M_B}
M= B \left(
       \begin{array}{cccc}
         \left(\frac{1+x}{1-x}\right)^{\lambda_1} & 0 & 0 & 0 \\
         0 & \left(\frac{1+x}{1-x}\right)^{\lambda_2} & 0 & 0 \\
         0 & 0 & \left(\frac{1+x}{1-x}\right)^{\lambda_3} & 0 \\
          0 & 0 & 0 & \left(\frac{1+x}{1-x}\right)^{\lambda_4} \\
       \end{array}
     \right) B^{T} .
\end{eqnarray}

The eigenvalues $\lambda_i$ can be found by comparing the singular behavior of the left and the right hand side of (\ref{M_B}) at $x\to \pm 1$.
Taking into account that only the matrix $N$ in $M$ is singular at $x \to \pm 1$, we find that $\lambda_1=1, \lambda_2=-1, \lambda_3=\lambda_4=0$.
Even more, if we write the matrix  $B$  in block form
\begin{eqnarray}
B= \left(
     \begin{array}{cc}
       B_1 & R \\
       L & B_2 \\
     \end{array}
   \right),
\end{eqnarray}
where $B_1$, $B_2$, $R$ and $L$ are $2\times 2$ matrices, from the singular behavior at  $x\to \pm 1$ we find

\begin{eqnarray}\label{S_Limit}
&&B_1 E_{\pm} {B_{1}}^{T}= \frac{1}{4} N_{\pm} ,\\
&&B_{1} E_{\pm} L^{T}=\frac{1}{4} N_{\pm}S_{\pm} ,\\
&&LE_{\pm}L^{T}= \frac{1}{4} {S_{\pm}}^{T} N_{\pm} S_{\pm} .
\end{eqnarray}
Here the matrices $E_{\pm}$, $N_{\pm}$ and $S_{\pm}$ are defined as follows

\begin{eqnarray}
&&E_{+}=\left(
        \begin{array}{cc}
          1 & 0 \\
          0 & 0 \\
        \end{array}
      \right), \;\;\;  E_{-}= \left(
                                \begin{array}{cc}
                                  0 & 0 \\
                                  0 & 1 \\
                                \end{array}
                              \right), \\
&&N_{\pm}= \lim_{x\to\pm 1} (1-x^2)N =e^{\sqrt{\frac{2}{3}}\varphi_{\pm} + \sigma_{\pm}}
 \left(\begin{array}{cc}
        4{\Psi^{\pm}}^2 & -2\Psi^{\pm} \\
         -2\Psi^{\pm} & 1 \\
        \end{array}
 \right), \\
&&S_{\pm} = \lim_{x\to \pm 1}S = \left(
                               \begin{array}{cc}
                                 2\Phi^{\pm}_1 & 2\Phi^{\pm}_{2} \\
                                 \chi^{\pm}_1 +  2\Phi^{\pm}_1 \Psi^{\pm} & \chi^{\pm}_{2} + 2\Phi^{\pm}_2 \Psi^{\pm} \\
                               \end{array}
                             \right).
\end{eqnarray}

In order to explore the regular part of  $M$ at $x\to 1$ we consider the matrix $(1-x)M$. Taking into account (\ref{S_Limit}) we find that at $x\to 1$
we have

\begin{eqnarray}
&&(1-x)N= \frac{1}{2}N_{+} +  (1-x) RR^{T} + \frac{1}{8}(1-x)^2 N_{-}  , \\
&&(1-x)NS = \frac{1}{2}N_{+}S_{+} +  (1-x)R B_{2}^{T} +  \frac{1}{8}(1-x)^2  N_{-}S_{-}   ,\\
&&(1-x)S^T N S + (1-x)Y= \frac{1}{2}S_{+}^{T} N_{+} S_{+} + (1-x)B_{2}B^T_{2} +  \frac{1}{8} (1-x)^2 S_{-}^T N_{-}S_{-} .
\end{eqnarray}

Using these relations after long but straightforward calculations we obtain

\begin{eqnarray}
\lim_{x\to 1} \frac{h}{1-x^2} \frac{H_{IJ}a^{I}_{+}a^{J}_{+}}{1-x^2}=
\frac{e^{-\sqrt{\frac{2}{3}} \,\varphi_{+} -\sigma_{+}}}{16} \left[s_{+}^{T}(a_{+}) - s_{-}^{T}(a_{+}) \right] N_{-} \left[s_{+}(a_{+}) - s_{-}(a_{+}) \right],
\end{eqnarray}
where

\begin{eqnarray}
s_{\pm}(a_{+})= S_{\pm} a_{+}= \left(
                               \begin{array}{cc}
                                 2\Phi^{\pm}_1 & 2\Phi^{\pm}_{2} \\
                                 \chi^{\pm}_1 +  2\Phi^{\pm}_1 \Psi^{\pm} & \chi^{\pm}_{2} + 2\Phi^{\pm}_2 \Psi^{\pm} \\
                               \end{array}
                             \right) \left(
                                       \begin{array}{c}
                                         a^{1}_{+} \\
                                         a^{2}_{+} \\
                                       \end{array}
                                     \right)=
                                     \left(
                                       \begin{array}{c}
                                         2\Phi^{\pm}_{I}a^{I}_{+} \\
                                        \chi^{\pm}_{I}a^{I}_{+} +   2\Phi^{\pm}_{I}a^{I}_{+}\Psi^{\pm} \\
                                       \end{array}
                                     \right).
\end{eqnarray}

By similar considerations one can show that

\begin{eqnarray}
\lim_{x\to -1} \frac{h}{1-x^2} \frac{H_{IJ}a^{I}_{-}a^{J}_{-}}{1-x^2}=
\frac{e^{-\sqrt{\frac{2}{3}} \,\varphi_{-} -\sigma_{-}}}{16} \left[s_{+}^{T}(a_{-}) - s_{-}^{T}(a_{-}) \right] N_{+} \left[s_{+}(a_{-}) - s_{-}(a_{-}) \right],
\end{eqnarray}
where

\begin{eqnarray}
s_{\pm}(a_{-})= S_{\pm} a_{-}= \left(
                               \begin{array}{cc}
                                 2\Phi^{\pm}_1 & 2\Phi^{\pm}_{2} \\
                                 \chi^{\pm}_1 +  2\Phi^{\pm}_1 \Psi^{\pm} & \chi^{\pm}_{2} + 2\Phi^{\pm}_2 \Psi^{\pm} \\
                               \end{array}
                             \right) \left(
                                       \begin{array}{c}
                                         a^{1}_{-} \\
                                         a^{2}_{-} \\
                                       \end{array}
                                     \right)=
                                     \left(
                                       \begin{array}{c}
                                         2\Phi^{\pm}_{I}a^{I}_{-} \\
                                        \chi^{\pm}_{I}a^{I}_{-} +   2\Phi^{\pm}_{I}a^{I}_{-}\Psi^{\pm} \\
                                       \end{array}
                                     \right).
\end{eqnarray}

The above results combined with (\ref{A_INEQ}) and (\ref{I_FUN}), when  $I_{*}[X^A]=0$  is taken into account, give the following inequality

\begin{eqnarray}
{\cal A}\ge 8\pi^2 \left(Z_{+}Z_{\,-}\right)^{1/4} ,
\end{eqnarray}
where

\begin{eqnarray}
Z_{+} = \frac{1}{16}[s_{+}^T(a_{+}) - s_{-}^T(a_{+})] \Sigma_{-} [s_{+}(a_{+}) - s_{-}(a_{+})], \\
Z_{\,-}= \frac{1}{16}[s_{+}^T(a_{-}) - s_{-}^T(a_{-})] \Sigma_{+} [s_{+}(a_{-}) - s_{-}(a_{-})],
\end{eqnarray}
and

\begin{eqnarray}
\Sigma_{\pm} = e^{-\sqrt{\frac{2}{3}} \,\varphi_{\pm} -\sigma_{\pm}} N_{\pm}= \left(
                 \begin{array}{cc}
                   4{\Psi^{\pm}}^2 & -2\Psi^{\pm} \\
                   -2\Psi^{\pm} & 1 \\
                 \end{array}
               \right) .
\end{eqnarray}

In order to express the inequality in more compact form  we should relate the potentials values at $x=\pm 1$ with the angular momenta, with  the charges
and with the magnetic fluxes. The full angular momenta $J_{I}$  associated with ${\cal B}$ are given by

\begin{eqnarray}
J_{I}= \frac{\pi}{4} \int_{{\cal {\hat B}}}i_{\eta_2}i_{\eta_1}\star d\eta_{I} -\frac{\pi}{2} \int_{\hat {\cal B}}(\Phi_{I}d\Psi - \Psi d\Phi_{I} )= \nonumber\\
\frac{\pi}{4} \int_{{\cal {\hat B}}}\omega_{I} -\frac{\pi}{2} \int_{\hat {\cal B}}(\Phi_{I}d\Psi - \Psi d\Phi_{I} )=\frac{\pi}{4}  \int_{{\cal {\hat B}}} d\chi_{I},
\end{eqnarray}
where the first integral is the contribution of the gravitational field while the second one reflects the contribution of the electromagnetic field.
The direct calculation gives the following expressions for $J_{I}$, namely

\begin{eqnarray}
J_{I} = \frac{\pi}{4}\left(\chi^{+} - \chi^{-}\right)= \frac{\pi}{2}\chi^{+}.
\end{eqnarray}
The electric charge is  given by
\begin{eqnarray}
Q = \frac{1}{2\pi^2} \int_{{\cal B}} e^{-2\alpha\varphi}\star F = 2\left(\Psi^{+} - \Psi^{-}\right)=4\Psi^{+}.
\end{eqnarray}

In this way we obtain

\begin{eqnarray}
{\cal A}\ge 8\pi \sqrt{|J_{+} + \frac{1}{8}Q \mathfrak{F}_{+}||J_{-} - \frac{1}{8}Q \mathfrak{F}_{-}|},
\end{eqnarray}
where

\begin{eqnarray}
J_{\pm}=J_{I}a^{I}_{\pm}, \;\;\; \;  \mathfrak{F}_{\pm}= 2\pi \left(
\Phi^{+}_{I} - \Phi^{-}_{I}\right) a^{I}_{\pm} .
\end{eqnarray}

The quantities $\mathfrak{F}_{\pm}$ can be interpreted as magnetic fluxes through appropriately defined 2-surfaces ${\cal D}_{\pm}$.
We define ${\cal D}_{\pm}$ in the following way. First we uplift the factor space interval ${\hat {\cal B}}=[-1,1]$ to a curve in the spacetime manifold
${\cal M}$ and then we act with the isometries generated by the Killing field $a^{I}_{\pm}\eta_{I}$. It is not difficult to see that  the so constructed
2-dimensional surfaces ${\cal D}_{\pm}$ have $S^2$-topology for ${\bf a}_{+}=\pm {\bf a}_{-}$ and  disk topology in the other cases. The magnetic
fluxes through  ${\cal D}_{\pm}$ are given by

\begin{eqnarray}
&&\mathfrak{F}_{\pm}= \int_{{\cal D}_{\pm}} F= 2\pi \int_{{\hat {\cal B}}} i_{a^{I}_{\pm}\eta_{I}}F= 2\pi a^{I}_{\pm} \int_{{\hat {\cal B}}} i_{\eta_{I}}F=2\pi a^{I}_{\pm} \int_{{\hat {\cal B}}} d\Phi_{I}=
2\pi a^{I}_{\pm} \int^{1}_{-1} d\Phi_{I} \nonumber \\&& = 2\pi a^{I}_{\pm} \left(\Phi^{+}_{I} - \Phi^{-}_{I}\right)
\end{eqnarray}
and obviously coincide with the previously defined quantities $\mathfrak{F}_{\pm}$. In the case when the topology of ${\cal D}_{\pm}$ is the spherical one,
the magnetic fluxes are in fact (up to sign) the magnetic (dipole) charge associated with ${\cal B}$.

Let us summarize the results of this section in the following

\medskip
\noindent

{\bf Theorem 1.} {\it Let ${\cal B}$ be a smooth stably outer marginally trapped surface in a spacetime satisfying
5D Einstein-Maxwell-dilaton equations with a dilaton coupling parameter $\alpha^2=\frac{8}{3}$  and having isometry group $U(1)^2$. If the dilaton potential is
non-negative, then the area of ${\cal B}$ satisfies the inequality }

\begin{eqnarray}
{\cal A}\ge 8\pi \sqrt{|J_{+} + \frac{1}{8}Q \mathfrak{F}_{+}||J_{-} - \frac{1}{8}Q \mathfrak{F}_{-}|}
\end{eqnarray}
{\it with $J_{\pm}=J_{I}a^{I}_{\pm }$, where $J_{I}$, $Q$, and $\mathfrak{F}_{\pm}$ are the angular momenta, the electric charge and
the magnetic fluxes associated with ${\cal B}$, respectively.
The equality is saturated only for the extremal stationary near horizon
geometry of the $\alpha^2=\frac{8}{3}$ Einstein-Maxwell-dilaton gravity  with
$V(\varphi)=0$.  }

\medskip
\noindent

\section{Horizon area-angular momenta-charge-magnetic fluxes inequality for dilaton coupling parameter $0\le \alpha^2\le \frac{8}{3}$ }
Finding a sharp lower bound for the horizon area for arbitrary dilaton coupling parameter is very difficult since the geodesic equations for arbitrary
$\alpha$ can not be integrated explicitly. Nevertheless an important estimate can be found for dilaton coupling parameter in the range
$0\le \alpha^2\le \frac{8}{3}$.  The inequality is given by the following

\medskip
\noindent

{\bf Theorem 2.} {\it Let ${\cal B}$ be a smooth stably outer marginally trapped surface in a spacetime satisfying
5D Einstein-Maxwell-dilaton equations with a dilaton coupling parameter $0\le\alpha^2\le\frac{8}{3}$  and having isometry group $U(1)^2$.
If the dilaton potential is non-negative, then the area of ${\cal B}$ satisfies the inequality }

\begin{eqnarray}
{\cal A}\ge 8\pi \sqrt{|J_{+} + \frac{1}{8}Q \mathfrak{F}_{+}||J_{-} - \frac{1}{8}Q \mathfrak{F}_{-}|}
\end{eqnarray}
{\it with $J_{\pm}=J_{I}a^{I}_{\pm }$, where $J_{I}$, $Q$, and $\mathfrak{F}_{\pm}$ are the angular momenta,
the electric charge and the magnetic fluxes associated with ${\cal B}$, respectively.
The equality is saturated  for the extremal stationary near horizon
geometry of the $\alpha^2=\frac{8}{3}$ Einstein-Maxwell-dilaton gravity  with
$V(\varphi)=0$.  }

\medskip
\noindent

{\bf Proof.} The proof is a direct generalization of the one in four dimensions \cite{Y1}. Let us first consider the case $0<\alpha^2\le \frac{8}{3}$ and define
the metric

\begin{eqnarray}
&&{\tilde G}_{AB}dX^AdX^B= \frac{1}{8} Tr\left(H^{-1}dH\right)^2 + \frac{1}{8}h^{-2}(dh)^2 + \nonumber \\
&&\frac{1}{4}h^{-1}H^{IJ}\left(d\chi_{I} +  2\Phi_{I}d\Psi - 2\Psi d\Phi_{I}\right) \left(d\chi_{J}
+ 2\Phi_{J}d\Psi - 2\Psi d\Phi_{J}\right) + \nonumber \\ &&  e^{-2\alpha\varphi }H^{IJ}d\Phi_{I}d\Phi_{J} + e^{2\alpha\varphi} h^{-1}(d\Psi)^2 +
\frac{3\alpha^2}{8}(d\varphi)^2
\end{eqnarray}

and the associated  functional

\begin{eqnarray}
{\tilde I}[X^A]= \int^{1}_{-1}\left[(1-x^2){\tilde G}_{AB}\frac{dX^A}{dx}\frac{dX^B}{dx} - \frac{1}{1-x^2}\right]dx
+ \frac{1}{4}x\ln\left[\frac{h}{1-x^2} \frac{H_{IJ}a^I(x)a^J(x)}{1-x^2}\right]|^{x=1}_{x=-1} .
\end{eqnarray}

It is not difficult to see that $I[X^A]\ge {\tilde I}[X^A]$ which gives
\begin{eqnarray}
{\cal A}\ge 8\pi^2 e^{{\tilde I}[X^A]}.
\end{eqnarray}

However, redefining the dilaton field ${\tilde \varphi}=\sqrt{\frac{3}{8}}\alpha \varphi$, we see that the functional ${\tilde I}[X^A]$ reduces to the
functional $I[X^A]$ for the critical coupling $\alpha^2=\frac{8}{3}$. Therefore we can conclude that

\begin{eqnarray}
{\cal A}\ge 8\pi \sqrt{|J_{+} + \frac{1}{8}Q \mathfrak{F}_{+}||J_{-} - \frac{1}{8}Q \mathfrak{F}_{-}|}
\end{eqnarray}
for every $\alpha$ in the range $0<\alpha^2\le \frac{8}{3}$. The continuity argument shows that the inequality also holds  for the Einstein-Maxwell case $\alpha=0$.

\section{Discussion}

In the present paper we derived inequalities between the area, the angular momenta, the electric charge  and
the magnetic fluxes for any smooth stably outer marginally trapped surface in 5D Einstein-Maxwell-dilaton gravity
with dilaton coupling parameter in the range $0\le \alpha^2 \le \frac{8}{3}$. In proving the inequalities we assumed that
the dilaton potential is non-negative and the spacetime is $U(1)^2$ axisymmetric but otherwise highly dynamical.
{\it It is worth mentioning that all of our results still hold even in the presence of matter with an axially symmetric energy momentum tensor
satisfying the dominant energy condition.}

Since the considerations in the present paper are entirely quasi-local, our results can be applied to stationary axisymmetric 
black holes in asymptotically flat and Kaluza-Klein spacetimes, as well as in spacetimes with de Sitter asymptotic.

The approach of the present paper can be easily extended to the case of 5D Einstein-Maxwell-Chern-Simons gravity with
with Chern-Simons coefficient $\lambda_{CS}$.

\medskip
\noindent

\medskip
\noindent

\noindent {\bf Acknowledgements:} This work was partially supported
by the Bulgarian National Science Fund under Grant DMU-03/6.


\begin{thebibliography}{99}


\bibitem{ADC} A. Acena, S. Dain and M.E. Gabach Clement, Class. Quant. Grav. {\bf 28}  105014
(2011); [arXiv:1012.2413[gr-qc]].

\bibitem{DR} S. Dain and M. Reiris. Phys. Rev. Lett. {\bf 107}, 051101 (2011); [arXiv:1102.5215[gr-qc]].

\bibitem{C} M.E. Gabach Clement, [arXiv:1102.3834[gr-qc]].

\bibitem{JRD} J. L. Jaramillo, M. Reiris and S. Dain,  Phys. Rev. {\bf D84}, 121503 (2011); [arXiv:1106.3743[gr-qc]].

\bibitem{DJR} S. Dain, J. L. Jaramillo and M. Reiris, Class. Quantum Grav. {29}, 035013 (2012); [arXiv:1109.5602[gr-qc]].

\bibitem{CJ} M.E. Gabach Clement and J.L. Jaramillo, [arXiv:1111.6248[gr-qc]].

\bibitem{S} W. Simon. Class. Quant. Grav. {\bf 29}, 062001 (2012);
[arXiv:1109.6140[gr-qc].]

\bibitem{CJR} M. E. Gabach Clement, J. L. Jaramillo and M. Reiris,
[arXiv:1207.6761[gr-qc]].

\bibitem{AHC} M. Ansorg, J. Hennig and C. Cederbaum. Gen. Rel. Grav. {\bf 43}, 1205 (2011); [arXiv:1005.3128[gr-qc]].

\bibitem{HAC} J. Hennig, M. Ansorg and C. Cederbaum. Class. Quantum Grav. {\bf 25} 162002 (2008);[arXiv:0805.4320[gr-qc]].


\bibitem{HCA} J. Hennig, C. Cederbaum and M. Ansorg, Commun. Math. Phys. {\bf 293}, 449
(2010); [arXiv:0812.2811[gr-qc]].

\bibitem{D} S. Dain. Class. Quant. Grav. {\bf 29}, 073001 (2012), [arXiv:1111.3615[gr-qc]].


\bibitem{CENS} P. Chrusciel, M. Eckstein, L. Nguyen and S. Szybka,
Class. Quant. Grav. {\bf 28}, 245017 (2011).

\bibitem{Mars} M. Mars, Class. Quant. Grav. {\bf 29}, 145019 (2012).

\bibitem{J} J. Jaramillo, Class. Quant. Grav. {\bf 29}, 177001
(2012).

\bibitem{Y1} S. Yazadjiev,  [arXiv:1210.4684v2[gr-qc]].

\bibitem{H} S. Hollands, Class. Quant. Grav. {\bf 29}, 065006 (2012); [arxiv:1110.5814[gr-qc]].


\bibitem{Y2} S. Yazadjiev, JHEP {\bf 1106},  083 (2011); [arXiv:1104.0378 [hep-th]].

\bibitem{Y3} S, Yazadjiev,  Phys. Rev. {\bf D73}, 104007 (2006); [arXiv:hep-th/0602116].

\bibitem{Y4} S. Yazadjiev,  JHEP {\bf 0607}, 036 (2006); [arXiv:hep-th/0604140].

\bibitem{Y5} S. Yazadjiev, Phys. Rev. {\bf D78}, 064032 (2008); [arXiv:0805.1600 [hep-th]].


\bibitem{HY1}
  S.~Hollands and S.~Yazadjiev, Commun. Math. Phys. {\bf 283}, 749 (2008) [arXiv:0707.2775 [gr-qc]].

\bibitem{HY2}
  S.~Hollands and S.~Yazadjiev, ``A uniqueness theorem for stationary Kaluza-Klein black holes``,
  Comm. Math. Phys. {\bf 32}, 631 (2011); [arXiv:0812.3036[gr-qc]]


\bibitem{HIW} S. Hollands, A. Ishibashi and R. Wald, Commun. Math. Phys. {\bf 271}, 699 (2007).

















\end{thebibliography}
\end{document}